The article has been published in *Science*, 385, 6714, 1241–1244 (2024)
Available under: https://www.science.org/doi/10.1126/science.adq4744

# Stoichiometric reconstruction of the $Al_2O_3$(0001) surface

Johanna I. Hütner[1]†, Andrea Conti[1]†, David Kugler[1], Florian Mittendorfer[1], Georg Kresse[2], Michael Schmid[1], Ulrike Diebold[1], Jan Balajka[1]*

[1]Institute of Applied Physics, TU Wien; Vienna, 1040, Austria.

[2]Faculty of Physics, University of Vienna; Vienna, 1090, Austria.

*Corresponding author. Email: jan.balajka@tuwien.ac.at

† These authors contributed equally

## Abstract:

Macroscopic properties of materials stem from fundamental atomic-scale details, yet for insulators, resolving surface structures remains a challenge. The basal (0001) plane of α-$Al_2O_3$ was imaged with noncontact atomic force microscopy with an atomically-defined tip apex. The surface forms a complex $(\sqrt{31} \times \sqrt{31})R\pm9°$ reconstruction. The lateral positions of the individual O and Al surface atoms come directly from experiment; how these connect to the underlying crystal bulk was determined based on computational modeling. Before the restructuring, the surface Al atoms assume an unfavorable, threefold planar coordination; the reconstruction allows a rehybridization with subsurface O that leads to a substantial energy gain. The reconstructed surface remains stoichiometric, $Al_2O_3$.

A key factor that governs all surface processes is the detailed arrangement and local coordination of atoms. While the surface structures of many materials have been established with atomic precision, wide–band-gap insulators remain largely uncharted territory due to the limited applicability of experimental techniques that rely on charged particles. A prototypical example is α-$Al_2O_3$ (corundum, sapphire), an insulator with high dielectric strength, mechanical hardness, chemical and thermal resistance, and excellent optical properties. Alumina is one of the most important materials in the ceramic industry, commonly used as an inert and non-reducible support for heterogeneous catalysis (*1*) and in microelectronics. Alumina is also an important model system for naturally occurring minerals, such as aluminosilicate clays (*2*), and is thought to be efficient in nucleating ice in clouds, affecting weather and climate (*3*).

The surface of alumina has long been known to rearrange when provided enough thermal energy; the first observation of a stable (√31 × √31)R±9° surface reconstruction was already made in the late 1960s (*4*, *5*). Since then, many research groups have tried to elucidate the details of alumina surfaces using diffraction (*6*–*8*), spectroscopy (*9*, *10*), noncontact atomic force microscopy (nc-AFM) (*11*–*13*), and computational modeling (*14*). A commonly accepted atomic model of the (√31 × √31)R±9° reconstruction consists of one or two metallic Al layers at the top (*6*, *7*, *13*), supposedly created by desorption of oxygen from the outermost layers upon high-temperature annealing in ultrahigh vacuum (UHV). According to X-ray diffraction experiments (*7*), the Al layers are rotated relative to the underlying bulk, which gives rise to a coincidence pattern with a (√31 × √31)R±9° periodicity. There is, however, no experimental evidence that the alleged metallic surface of this perfect insulator is conducting (*15*). Moreover, a surface with metallic Al is expected to be unstable: the Al evaporation rate is high at the temperatures where the reconstruction forms (>1000 °C), and an Al surface should be reactive in air, in conflict with experimental observations (*6*–*8*).

This work used noncontact atomic force microscopy (nc-AFM) based on the stiff qPlus force sensor (*16*) to overcome the inherent difficulty of imaging the surface of a wide-gap insulator with atomic resolution. Crucially, the atomically controlled tip apex (*17*) provided chemical identification of individual surface atoms via variations in image contrast. The nc-AFM images were then used to extract the atomic positions of the α-$Al_2O_3$(0001)-(√31 × √31)R±9° surface. The determination of the subsurface layers was assisted by density functional theory (DFT) combined with machine-learned force fields (MLFF). Interestingly, the composition of the extensively reconstructed layer is very close to that of the bulk, while the surface energy is much lowered compared to a simple bulk truncation. The reconstruction is driven by changing the coordination of surface cations to configurations that allow bonding to subsurface O atoms.

**Noncontact AFM resolves individual Al and O atoms of the reconstructed surface**

The α-$Al_2O_3$(0001) samples formed the (√31 × √31)R±9° reconstruction upon annealing in UHV or $10^{-6}$ mbar oxygen to 1000 °C and no further changes were observed at higher temperatures (up to 1300 °C). Annealing above 1000 °C removed any surface contamination, as evidenced by X-ray photoelectron spectroscopy (XPS, Fig. S1). Nc-AFM images of the √31-reconstructed surface (Fig. 1) demonstrate structural diversity of the reconstruction unit cell, including characteristic triangular (green) and hexagonal (yellow) features, observed previously by Lauritsen *et al.* (*13*). Some triangular regions contain a bright defect in the center (dashed circles), in the following identified as a surface Al vacancy. The √31 periodicity and the 9° rotational alignment from the bulk lattice are evident in reciprocal space, in the fast Fourier transform (FFT) of the AFM image and the low-energy electron diffraction (LEED) pattern of the reconstructed surface, both shown in Fig. 1B. LEED simultaneously detects both rotational domains of the reconstructed surface leading to a superposition of two diffraction patterns rotated by ±9°.

Individual surface atoms can be identified chemically when the apex of the AFM tip is controlled and known. Using a CuOx-terminated tip, introduced by Mönig *et al.* (*17*, *18*), results in a drastic change in image contrast compared to a more commonly used Cu-terminated tip, as shown in Fig. S2. For an O-terminated tip, repulsion from the surface O locally increases the oscillation frequency, displayed as bright in Fig. 1, whereas the attraction between the tip and the surface Al shifts the frequency down (dark in the image). The lateral positions of individual surface Al and O atoms can thus be determined based on the local frequency variations (Fig. 1D). The atomically resolved nc-AFM image in Fig. 1C reveals a variety of local arrangements. Near the corners of the unit cell, the bright O atoms form a honeycomb pattern. Surface O atoms surrounding Al atoms in distorted squares and triangles are shown in Fig. 1D on top of a symmetry-averaged image to reduce noise and tip asymmetries (see Fig. S3).

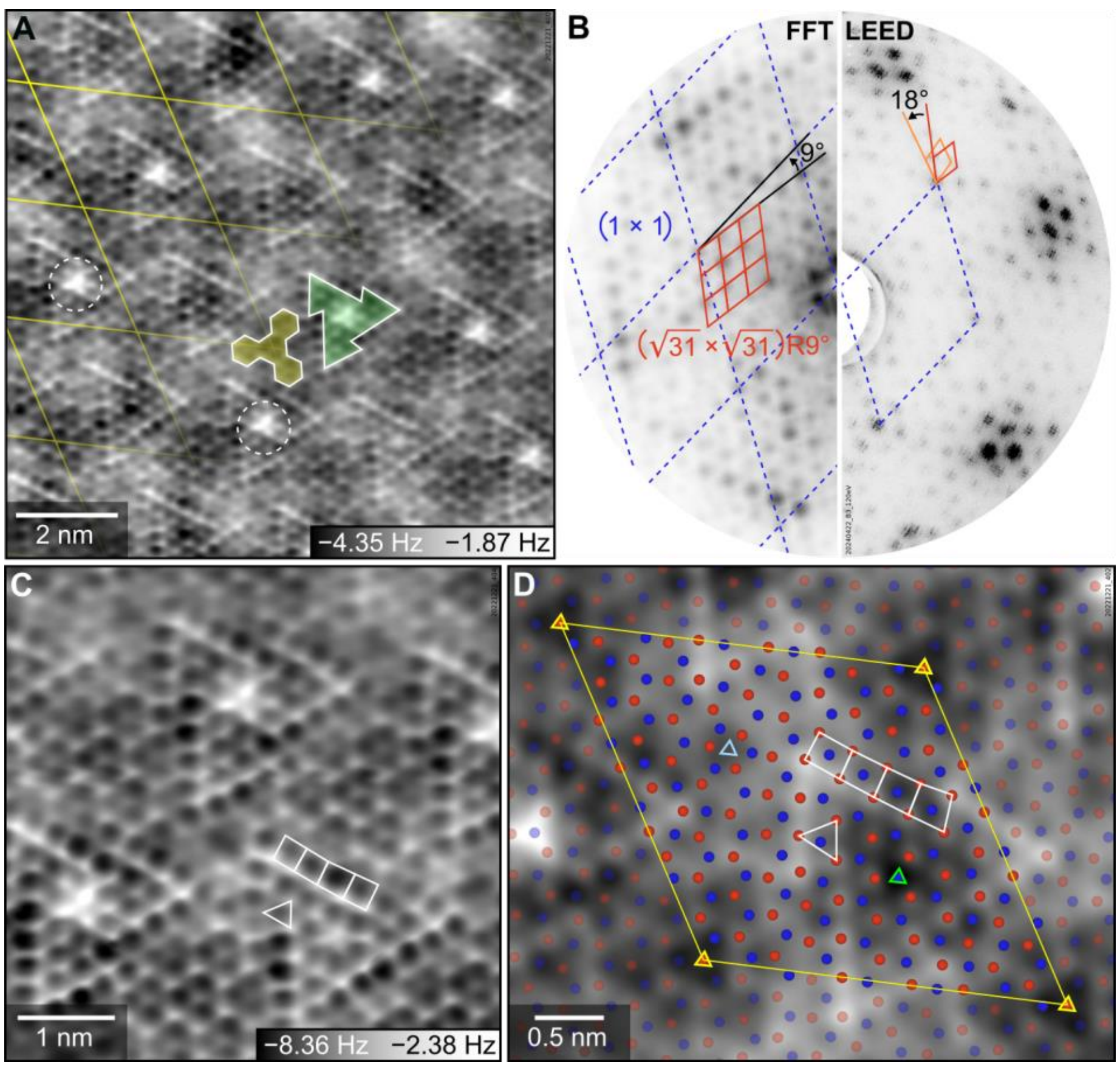

**Fig. 1. The (√31 × √31)R±9° reconstructed surface of $Al_2O_3$(0001).** (**A**, **C**) Noncontact-AFM images acquired at 4.7 K with a CuOx terminated tip at constant height. Oscillation amplitudes of 500 pm (A) and 300 pm (C) and a sample bias of −2.7 V were applied. Square and triangular arrangements of the surface oxygen atoms are marked by white lines, characteristic features are marked in yellow and green, and two defects (surface Al vacancies) are labeled by white dashed circles. A yellow grid indicates the √31 reconstruction lattice. (**B**) FFT of the AFM image in (A) and LEED pattern recorded at 120 eV. A blue dashed grid marks the bulk (1 × 1) reciprocal lattice.

The rotational domains (±9°) of the reconstruction lattice are marked in red and orange. (**D**) Symmetry-averaged nc-AFM image with atomic positions derived from the maxima and minima in the bandpass-filtered image in Fig. S3 (O red, Al blue). Colored triangles indicate the threefold symmetry centers.

## DFT calculations accelerated with machine learning identified the stable structure

The experimental nc-AFM images provide the lateral positions of the surface $Al_s$ and $O_s$ atoms. To create a complete three-dimensional model of the $Al_2O_3(0001)$-($\sqrt{31} \times \sqrt{31}$)R±9° reconstruction, the interface layers that connect the experimentally accessible surface to the bulk corundum structure need to be determined. The squares and triangles of surface $O_s$ atoms with an $Al_s$ at the center (Fig. 1) are the same structural motifs as in ultrathin alumina films (*19*, *20*). All these structures exhibit a distorted hexagonal $Al_s$ lattice with Al–Al distances of ≈3.04 Å. Based on this similarity, the building blocks of the √31 reconstruction were identified as $Al_s$ in tetrahedral or pyramidal coordination to the surrounding $O_s$ atoms in the surface and one interface $O_i$ atom located below each $Al_s$ (inset in Fig. 2B). These interfacial $O_i$ atoms form the apices of downward pointing tetrahedra and pyramids (truncated octahedra). The number of $O_i$ atoms is thus identical to the number of $Al_s$ atoms (76 per √31 unit cell) determined from AFM. Compared to corundum, with 93 O atoms per layer, the interfacial oxygen layer is rather defective (Fig. 2D). The $O_i$ atoms are expected to bind to two or three interfacial $Al_i$ atoms below to complete their coordination. Assuming a close-to-stoichiometric surface, the number of $Al_i$ atoms should be around 62–63 per √31 unit cell. In the corners of the reconstructed unit cell, the bright surface $O_s$ atoms form an interconnected honeycomb pattern (yellow in Fig. 1A), suggesting a similar arrangement of the $Al_i$ atoms below (*20*). The similarity with the honeycomb structure of Al atoms in bulk corundum indicates bulk-like stacking in these areas. The part of the unit cell marked by the green triangular area in Fig. 1A exhibits a dense hexagonal arrangement of octahedrally-coordinated $Al_i$ atoms with no honeycomb-like brightness modulations. Based on these considerations, initial structure models were created manually and optimized computationally, varying the number of Al and O atoms at the interface to account for the yet unknown stoichiometry. DFT calculations were combined with machine-learned force fields to accelerate the search as described in the Supplementary Materials. More than 100 initial structures were obtained by simulated annealing, and ~50 more were created manually and relaxed to test their viability.

The model of the √31 reconstruction in Fig. 2 fulfills two critical structure search criteria simultaneously: the lowest surface energy in the range of relevant oxygen chemical potentials and the best agreement between experimental and simulated AFM images. Consistent with XPS data (Fig. S4), the reconstructed surface terminates with a layer of $O_s$ that protrudes above the first layer of $Al_s$ (Fig. 2A). The top view (Fig. 2B) shows the relaxed arrangement of surface atoms, with only small root-mean-square deviations (0.15 Å and 0.21 Å for $Al_s$ and $O_s$, respectively) from the positions obtained by nc-AFM. The distorted hexagonal $Al_s$ lattice is apparent in Fig. 2C, where the $O_s$ atoms were removed. The interface layer (Fig. 2D) consists of two main regions with different arrangements of $O_i$ and $Al_i$ atoms. Near the corners of the √31 unit cell, the $Al_i$ atoms are close to their bulk positions and form a honeycomb pattern as in the corundum structure, and the $O_i$ atoms form a rather open structure. In the triangular region (near the symmetry center marked by a green triangle in Fig. 2D), the $O_i$ and $Al_i$ atoms form a denser, hexagonal lattice. Notably, the Al atoms in this region are stacked differently from the usual on-top stacking in corundum (dashed green lines in Fig. 2A), creating a stacking fault. These two regions dominate the contrast of experimental nc-AFM images in Fig. 1, honeycomb (yellow) and triangular (green, stacking fault).

The complete structure model, including two subsurface layers, is shown in Fig. S5. The troughs in the interfacial structure between the stacking-fault and honeycomb regions are consistent with fewer oxygen atoms compared to bulk corundum.

The high local density of positively charged Al within the stacking-fault region (Fig. 2B and D) has two consequences. First, the most favorable model has an Al vacancy $V_{Al}$ below the stacking fault in the bulk layer just below the interface (see Fig. 2 and Fig. S5). Second, high-temperature MD simulations suggest that a surface Al atom can be removed from the center of the triangular regions. The preference for creating an $Al_s$ vacancy at the (symmetry) center of the stacking-fault region is confirmed by the calculations shown in Fig. S6. Upon removal, the three surrounding (undercoordinated) O atoms relax upward, causing the bright triangular appearance frequently occurring in the nc-AFM images, see Fig. 1 and Fig. S7.

The structure with a surface Al vacancy at the symmetry center is nearly isoenergetic with the one in Fig. 2 in the range of experimental conditions (Fig. 3C). Coexistence of both defective and non-defective unit cells is thus expected, in agreement with experimental nc-AFM images (Fig. 1). The structure in Fig. 2 is only slightly reduced, with a deficit of one oxygen atom per √31 reconstruction unit cell (√31 O-poor) compared to stoichiometric $Al_2O_3$, while the structure with the additional $Al_s$ vacancy has a surplus of ½ oxygen atom per √31 reconstruction unit cell (√31 O-rich). Thus, the coexistence of the two structures at the surface is very close to stoichiometry, countering the idea that the √31 reconstruction should be a reduced surface. In addition to the structure in Fig. 2 and its variant with an Al vacancy, the nc-AFM images reveal slight structural irregularity at the border of the stacking-fault region in some unit cells (*e.g.*, see the arrow in Fig. S2B). The deviations from the usual appearance indicate variability at the interface to possibly accommodate any remaining nonstoichiometry while the surface layer remains unchanged.

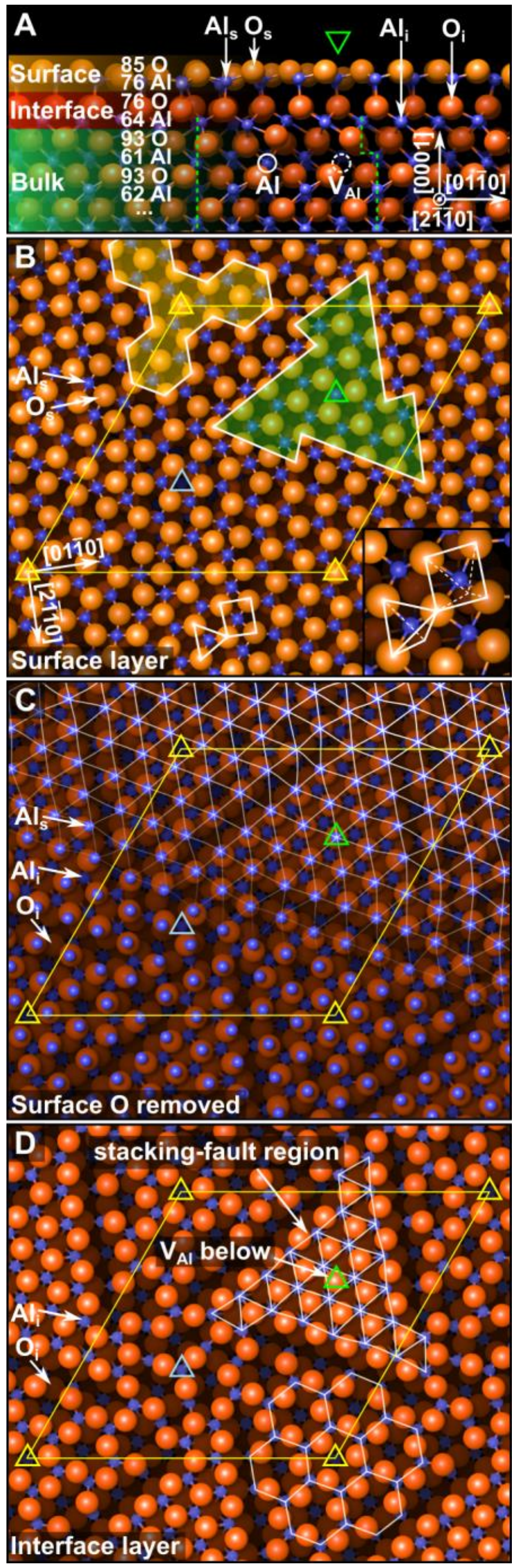

**Fig. 2. Structure model of the $Al_2O_3$(0001)-($\sqrt{31} \times \sqrt{31}$)R±9° surface.** (**A**) Section view and (**B-D**) top views with the $O_s$ removed in (C) and all surface atoms removed in (D). The numbers of Al and O atoms per ($\sqrt{31} \times \sqrt{31}$)R±9° unit cell are given in each layer in (A). The $\sqrt{31}$ unit cell is indicated by yellow rhombi, and the centers of symmetry are marked by colored triangles. Characteristic square, triangular, and honeycomb motifs observed in experimental images are indicated. The white triangular grid in (D) marks the stacking-fault region of the $Al_i$ atoms.

### *Simulated AFM reproduces experimental contrast*

For a direct comparison with the experimental nc-AFM images, simulated AFM images of the structure models were generated. The simulated image of the stable structure in Fig. 2 shows a remarkable correspondence with the experimental image, as shown in Fig. 3A and B. To demonstrate the sensitivity of this method, a simulated AFM image of the second-best structure with the same stoichiometry is presented in Fig. S8. The differences in the structures are relatively minor and limited to the boundary of the stacking fault region in the interface layer, yet the visual agreement with the experimental image is noticeably worse. The sensitivity of the simulated AFM images is not limited to high-contrast areas, but it equally applies to the low-contrast region near the light blue triangle in Figs. 2 and 3. AFM simulations thus serve as a valuable tool for verifying the structural models.

### *The √31 reconstructed surface is strongly favored over other structures*

The relative stabilities of selected structure models are compared in an *ab initio* phase diagram in Fig. 3C. The models with and without a surface Al vacancy discussed above (red and yellow lines in Fig. 3C, respectively) exhibit the lowest surface energy over a wide range of oxygen chemical potentials. The previously proposed Al adlayer model (*13*) (solid black line) is extremely unfavorable except at unrealistically low oxygen chemical potentials (< −5 eV). Compared to the √31 reconstruction, also the unreconstructed, relaxed (1 × 1) surface is very unfavorable (almost 50 % higher in energy).

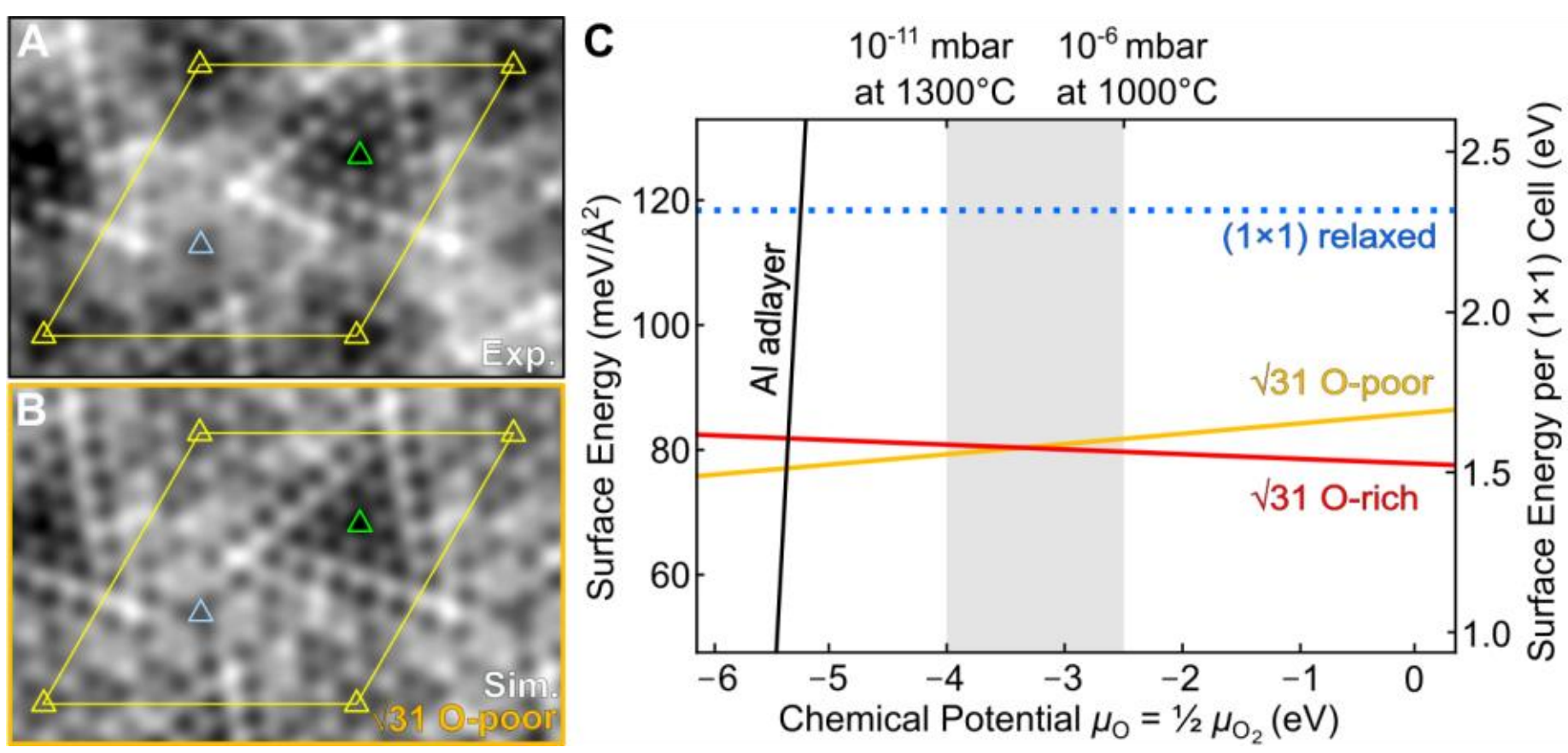

**Fig. 3. Stable $Al_2O_3$(0001) termination matches experimental AFM.** (**A**) Experimental and (**B**) simulated AFM images of the lowest energy model in Fig. 2. (**C**) *Ab initio* phase diagram. The surface energies are plotted as a function of oxygen chemical potential; the gray stripe highlights the range of experimental conditions. In this range, the O-poor model in Fig. 2 and its O-rich version (with a surface Al vacancy) are nearly isoenergetic.

### *The structure model agrees with existing experimental data*

The structure of the reconstructed $Al_2O_3$(0001) surface determined in this work is distinctly different from previously suggested models. However, it agrees with existing experimental data. Our AFM images reproduced previous measurements (*11*, *13*), albeit with higher resolution thanks to the stiff qPlus nc-AFM sensor and chemical sensitivity based on the controlled tip apex. The new model, with two reconstructed Al and O atomic layers (Fig. 2), corroborates the presence of two distorted hexagonal Al planes with interatomic distances of 3.03 Å atop the corundum bulk, established by Renaud, *et al.*, via surface X-ray diffraction (SXRD) experiments (*7*). Similarly, the model proposed here contains a layer with a nearly hexagonal lattice of surface $Al_s$ atoms (Fig. 2C) with an almost identical average Al–Al in-plane distance of 3.04 Å and a second Al plane with a similar lattice in the stacking-fault region (Fig. 2D). Complementing X-ray diffraction, noncontact AFM, particularly with an oxygen-terminated tip, is suited to detect oxygen atoms at the surface and directly visualize the intricate local atomic configurations.

**The reconstruction is stabilized by increasing Al coordination**

Why does such a restructuring occur on the basal plane of a seemingly simple binary oxide? The bulk lattice of corundum is composed of hexagonal close-packed, (0001)-oriented oxygen layers separated by pairs of laterally shifted hexagonal Al planes, resulting in a stack of stoichiometric $(Al–O_3–Al)_n$ units. Each Al is coordinated to three O atoms from the layers above and below (Fig. S9A). A cut along (0001) between the Al planes of two adjacent $Al–O_3–Al$ units creates a non-polar surface, the standard model of the (1 × 1) termination. DFT calculations show that the undercoordinated terminal Al atoms on the (1 × 1) surface strongly relax inwards by 0.7 Å, resulting in an almost co-planar geometry (Al is only 0.14 Å above the O layer; Fig S9B). In the relaxed geometry, the short Al–O bond length of 1.69 Å (cf. Fig. S10) leads to compressive stress in the surface layer, with an energy cost of ≈0.1 eV per (1 × 1) cell. The strain energy alone cannot explain the difference in surface energies of 0.75 eV per (1 × 1) cell between the relaxed (1 × 1) surface and the reconstruction (Fig. 3C). The nearly threefold-planar coordination of the topmost Al of the (1 × 1) termination implies that their $p_z$ orbitals cannot bind to any oxygen neighbor, as also indicated by the projected density of states (Fig. S11). The energy cost for a non-bonding Al $p_z$ orbital can be estimated by comparison with threefold-planar aluminum trihalide molecules and their dimers: These molecules achieve a distorted tetrahedral coordination by dimerization (inset in Fig. S12). The energy gain upon dimerization is 1.11 eV per Al for $AlF_3 \rightarrow Al_2F_6$ and 0.64 eV for $AlCl_3 \rightarrow Al_2Cl_6$ (*21*). As Al prefers to bind to electronegative elements and the electronegativity of oxygen is between that of F and Cl, we estimate that the energy cost of a non-bonding Al $p_z$ is around 0.8 eV relative to a coordination where all Al 3*s* and 3*p* orbitals take part in bonding (Fig. S12). Converting the standard model of the (1 × 1) surface to any structure where the surface aluminum atoms have at least tetrahedral coordination (such as the √31 reconstruction) avoids one non-bonding Al $p_z$ orbital per (1 × 1) cell. The estimated ≈0.8 eV energy gain per (1 × 1) cell explains the magnitude of the surface energy difference upon reconstruction. The tetrahedral or pyramidal coordination of the surface Al atoms requires an in-plane expansion of the lattice at the surface. The complex √31 reconstruction is the result of fitting this expanded surface layer onto the corundum bulk structure.

The $Al_2O_3(0001)$ surface is substantially stabilized by adopting the √31 reconstruction, and recent experimental evidence (*8*) shows this structure is stable at all attainable oxygen chemical potentials. The basal planes of other corundum oxides, such as $Cr_2O_3$ (*22*), $V_2O_3$ (*23*), and $Fe_2O_3$ (*24*), show a tendency to reconstruct, and a striking resemblance is evident between reconstructed alumina and hematite $Fe_2O_3(0001)$ (*25*). This similarity suggests a unified mechanism causing the reconstruction by avoiding undercoordinated metal cations and creating an O-terminated surface.

## Acknowledgments:

The authors thank H. Mönig and P. Wiesener for sharing their expertise with CuOx-terminated tips. The computational results presented have been achieved using the Vienna Scientific Cluster (VSC). The work was supported by European Research Council grant 883395, and by Austrian Science Fund (FWF) grant SFB-F81, doi: 10.55776/F81.

**Author contributions:**

Conceptualization: JB, MS, UD

Methodology: JB, MS, FM, GK

Investigation: JIH, AC, DK, JB, MS

Visualization: JIH, AC, JB, MS

Funding acquisition: UD, GK

Project administration: JB, UD, MS

Supervision: JB, MS, FM, UD

Writing – original draft: JB, JIH, AC, MS

Writing – review & editing: JB, MS, UD, FM, GK

The authors declare no competing interests.

Supplementary Materials for

# Stoichiometric reconstruction of the $Al_2O_3(0001)$ surface

Johanna I. Hütner[1]†, Andrea Conti[1]†, David Kugler[1], Florian Mittendorfer[1], Georg Kresse[2], Michael Schmid[1], Ulrike Diebold[1], Jan Balajka[1]*

[1]Institute of Applied Physics, TU Wien; Vienna, 1040, Austria.

[2]Faculty of Physics, University of Vienna; Vienna, 1090, Austria.

*Corresponding author. Email: jan.balajka@tuwien.ac.at

† These authors contributed equally

## Materials and Methods

### Sample preparation

Experiments were performed on polished α-$Al_2O_3(0001)$ single crystals from Crystec GmbH, cleaned by repeated cycles of sonication in pH-neutral detergent (3% Extran MA 02 from Merck, heated at 40 °C) and rinsing with ultrapure $H_2O$ (Milli-Q, Millipore, 18.2 MΩ cm, <3 ppb total organic carbon) until no polishing residues were detected with ambient AFM (Agilent 5500). Cleaned samples were annealed in a tube furnace in air at 1100 °C for 10 h to obtain a smooth surface with wide atomic terraces and straight step-edges and afterwards mounted on Omicron-type sample plates, made from Ta, using Ta retaining wires spot welded to the sample plate. The samples were then annealed using a resistively heated W filament in the preparation chamber (base pressure < $2\times10^{-10}$ mbar) of an ultrahigh vacuum (UHV) system. Annealing temperatures vs. heating power were calibrated with a K-type thermocouple mounted on an empty sample plate at the sample position. After annealing, samples were transferred to the analysis chamber (base pressure of < $1\times10^{-11}$ mbar) equipped with an Omicron qPlus LT STM/AFM cooled to 4.7 K by a He bath cryostat and a cryogenic differential amplifier (*1*).

### Noncontact atomic force microscopy

Noncontact atomic force microscopy (nc-AFM) imaging was conducted in constant-height mode using a qPlus sensor (*2*) ($f_0$ = 40.2 kHz, $k$ = 2000–3500 N/m, Q = 22000) with an electrochemically etched tungsten tip attached to the oscillating prong. Before imaging, the tip was treated with field emission and self-sputtering with $Ar^+$ ions, as described in ref. (*3*), and further conditioned by voltage pulses on clean Au(100) and Cu(110) surfaces. Oxygen-terminated tips were prepared following the procedure described in ref. (*4*). This procedure includes voltage pulses and tip indentation on a Cu(110) single crystal partially covered by an oxygen-induced surface reconstruction formed upon a brief $O_2$ exposure ($2\times10^{-8}$ mbar for 15 s) at elevated temperature (250 °C). The characteristic contrast evolution with varying distances between the tip and the partially oxidized Cu(110) was used to verify a stable oxygen tip termination (Fig. S2). The nc-AFM images were recorded with sample bias voltages adjusted between −1.3 V and −2.7 V to minimize the local contact potential difference (LCPD). Raw images were corrected for distortions due to piezo creep (*5*) and filtered in the Fourier domain to remove mechanical and electrical noise.

### X-ray photoelectron spectroscopy and low-energy electron diffraction

X-ray photoelectron spectroscopy (XPS) data were acquired in the preparation chamber with a non-monochromatized Al Kα source (SPECS XR 50) and a SPECS Phoibos 100 hemispherical analyzer at 20 eV and 60 eV pass energy, in normal emission (0°) and at 70° from the surface normal to increase surface sensitivity. A linear Shirley background was subtracted from the O 1*s* spectrum collected at 0° before fitting two components, assuming asymmetric Lorentzian line shapes to account for spectrum asymmetry. The binding energy of the main component, identified as a contribution from bulk $Al_2O_3$, was adjusted to 531.0 eV, and the same offset (7.6 eV) was applied to the entire spectrum to compensate for sample charging. The O 1*s* and Al 2*p* spectra were normalized to the integrated area of the O 1*s* spectrum at the respective emission angle. Low-energy electron diffraction (LEED) images were recorded in the preparation chamber using SPECS ErLEED 150. Flat field images (LEED of a polycrystalline sample holder) and dark frames (screen voltage off) were used to correct for screen inhomogeneities (*6*).

### *Ab initio* calculations

Density functional theory (DFT) calculations were performed using the projector-augmented wave (PAW) method (*7*, *8*) as implemented in the Vienna Ab-initio Simulation Package (VASP) (*9*, *10*), using the standard oxygen (O) and aluminum (Al) PAW potentials, the semilocal Perdew-Burke-Ernzerhof (PBE) (*10*) exchange-correlation functional for training the force fields, and the $r^2$SCAN metaGGA (*11*) functional for the determination of the ground state structures and energies. These functionals describe well the structural properties of bulk α-$Al_2O_3$ (lattice constants *a* and *c* deviate from the experimental values (*12*) by +0.9% for PBE, and +0.1% and −0.2%, respectively, for $r^2$SCAN). The bulk structure was optimized with a cutoff energy of 800 eV, and a k-point mesh of 3 × 3 × 1 was used to integrate the Brillouin zone. The surface calculations were performed with a lower cutoff energy of 500 eV and only evaluating a single k-point (Γ) for the (√31 × √31)R9° cell. For geometry optimizations, the conjugate gradient method was used, and the structures were relaxed until the residual forces on the atoms were less than 0.01 eV/Å and an energy convergence of $10^{-6}$ eV was achieved for the final configuration.

The chemical potential of oxygen in the *ab initio* phase diagram in Fig. 3 is defined as

$$\mu_O(T,p) = \frac{1}{2}\mu_{O_2}(T,p) = \mu_O(T,p^o) + \frac{1}{2}kT\ln\left(\frac{p}{p^o}\right),$$

which provides the temperature and pressure dependence, if the temperature dependence of $\mu_O(T,p^o)$ at a particular pressure, $p^o$, is known (*13*). The reference state of $\mu_O(T,p)$ was chosen as half of the total energy in a spin-polarized calculation for an isolated $O_2$ molecule in the gas phase.

### Computational modeling of the √31 reconstruction

The reconstruction models have a (√31 × √31)R−9° unit cell (in-plane dimensions: 26.5 Å × 26.5 Å). The top (surface) layer of the slab was constructed based on in-plane coordinates from the experimental nc-AFM images. Below, an intermediate interface layer was placed, which connects the surface with two bulk corundum layers underneath. The bottom of the asymmetric slab was oxygen-terminated and saturated with hydrogen to avoid a polar surface. Periodically repeating slabs were separated by a 20 Å vacuum spacing.

The surface, interface, and the first bulk layer were allowed to relax while the bottom bulk layer was fixed. Consistency checks on some configurations were performed using the metaGGA $r^2$SCAN functional to ensure the reliability of the PBE functional; $r^2$SCAN was also used to obtain the energies and structures of the final models. Machine-learned force fields (MLFF, implemented in VASP 6.4.1) were employed to accelerate the search through many possible structures. The force field was initially trained on smaller cells, containing the structures found in the reconstructed surface layer in the following order: bulk $\alpha$-$Al_2O_3$, bulk $\gamma$-$Al_2O_3$, (1 × 1) stoichiometric slabs, (√3 × √3) reconstructed slabs with different stoichiometries, and finally on (√31 × √31) reconstructed slabs with different stoichiometries. All structures were trained by molecular dynamics (MD) simulations starting at 1500 K; the temperature was gradually increased and maintained at 2300 K (near the melting point). This procedure generated a well-trained force field with more than 3400 local configurations, which was then used to perform MD simulations at high temperatures and subsequent geometry relaxations (simulated annealing) of different candidate structures in prediction-only mode (*i.e.*, without additional DFT calculations). Finally, each configuration found using the force field was relaxed in a standard DFT calculation. This approach significantly reduced the time required for structure optimization using standard *ab initio* calculations.

**AFM simulations**

Nc-AFM images of the DFT-relaxed models were simulated with the Probe Particle Model (*14*). The parameters of a $CuO_x$ tip ($k_{x,y}$ = 161.9 N/m, $k_z$ = 271.1 N/m, effective tip charge −0.05 e) were taken from ref. (*4*); note that with values larger than 5 N/m, the tip's stiffness had only a minor influence on the appearance of the simulated images. The oscillation amplitude of 300 pm was chosen to match the experiment. Since the exact height of the tip in the experiment is not known, the tip height was varied in the simulation and the height that yielded the best visual agreement between experiment and simulation was selected.

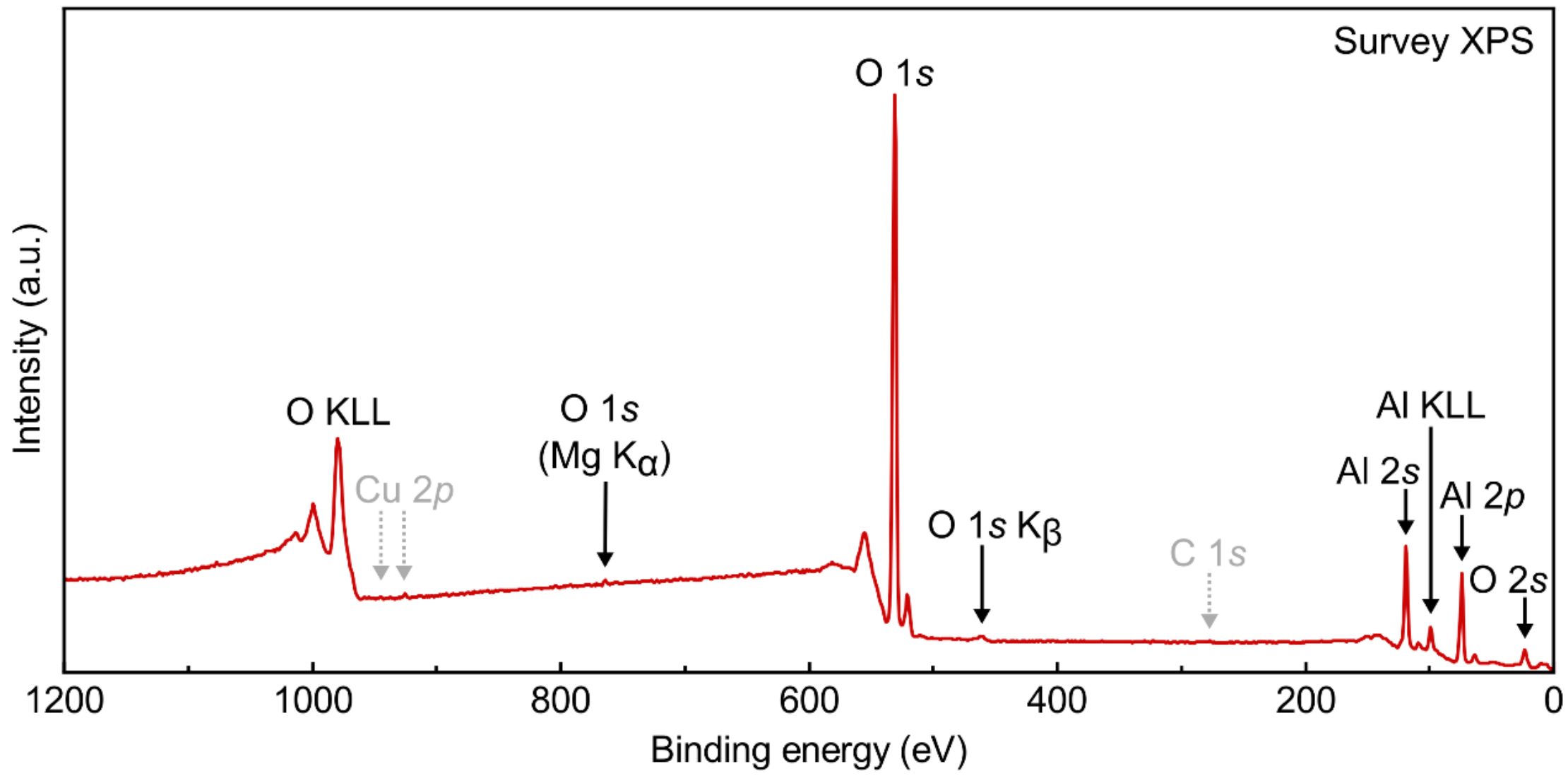

**Fig. S1. Survey XPS of the (√31 × √31)R±9° reconstructed $Al_2O_3$(0001) surface annealed in UHV.**

The spectrum was recorded at 70° from the surface normal at pass energy of 60 eV using Al Kα X-rays. All transitions originating from the sample are labeled in black. The spectrum confirms that no contamination was detected on the sample. Transitions indicated in gray are caused by a measurement artifact. These peaks do not shift in energy in response to an applied voltage between the sample and the ground and, therefore, do not originate from the sample. The spectrum was corrected for sample charging by a rigid shift, setting the O 1*s* binding energy to 531.0 eV.

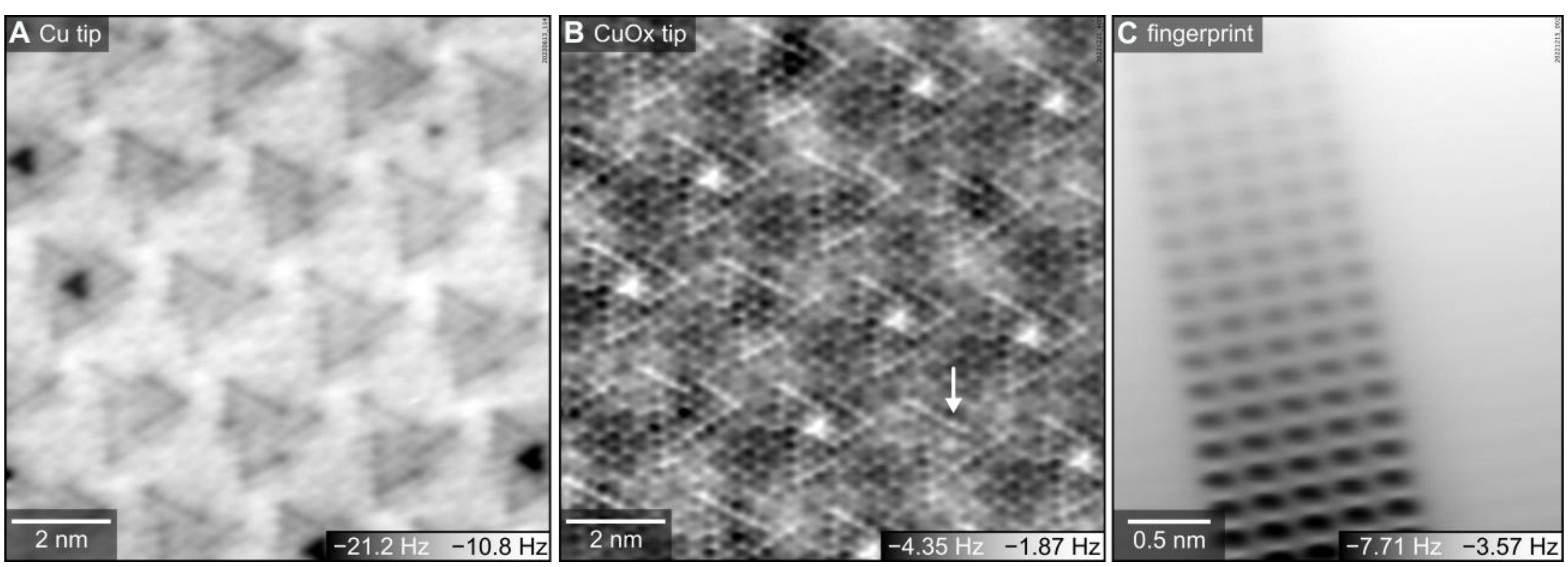

**Fig. S2. Noncontact-AFM images with Cu and CuOx tips and the AFM "fingerprint" image of Cu(110)-(1 × 2)-O to verify the tip termination.**

(**A,B**) Nc-AFM images of the √31 reconstructed $Al_2O_3$(0001) surface acquired at 4.7 K with (A) a Cu-terminated tip, and (B) CuOx-terminated tip, at constant height. The following oscillation amplitudes and sample bias were applied: (A) 80 pm, −0.8 V; (B) 500 pm, −2.7 V, (C) 500 pm, +0.001 V. The arrow in (B) marks a region that differs from the usual appearance. As discussed in the main text, the change in appearance is caused by structural variability at the stacking fault boundary in the interface layer. (**C**) Nc-AFM image of a (1 × 2) oxygen superstructure on Cu(110) acquired to verify the CuOx tip termination as described in ref. (*4*). The tip-surface separation is decreased during a downward scan to visualize the height-dependent contrast and distinguish the repulsive (bright) surface O atoms from the attractive (dark) Cu atoms.

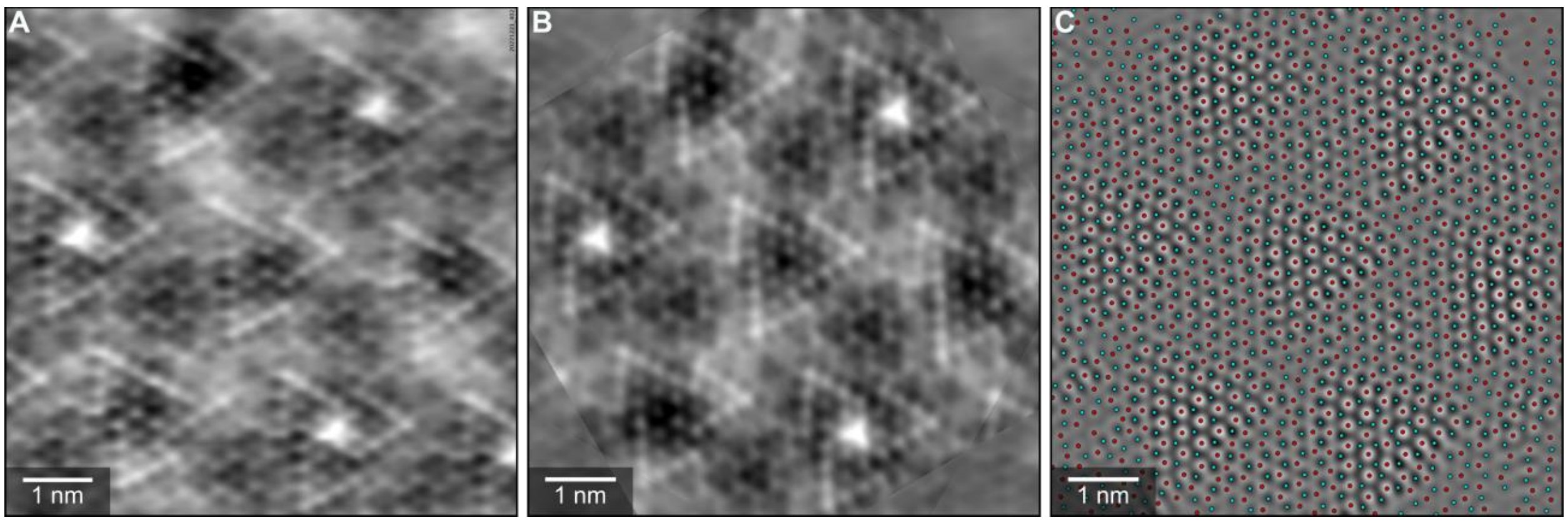

**Fig. S3. Determination of O and Al atomic positions from nc-AFM image.**
(**A**) Undistorted, FFT-filtered, and background-subtracted nc-AFM image. (**B**) Symmetry-averaged image to remove any tip asymmetry and reduce experimental noise by averaging the image in panel (A) with its copies rotated by 120° and 240° around the three-fold symmetry center. (**C**) Bandpass-filtered image in panel (B) for an automatized extraction of the local maxima and minima, which correspond to the O and Al positions, respectively. The atomic positions are overlaid on the (unfiltered) symmetry-averaged image in Fig. 1D in the main text.

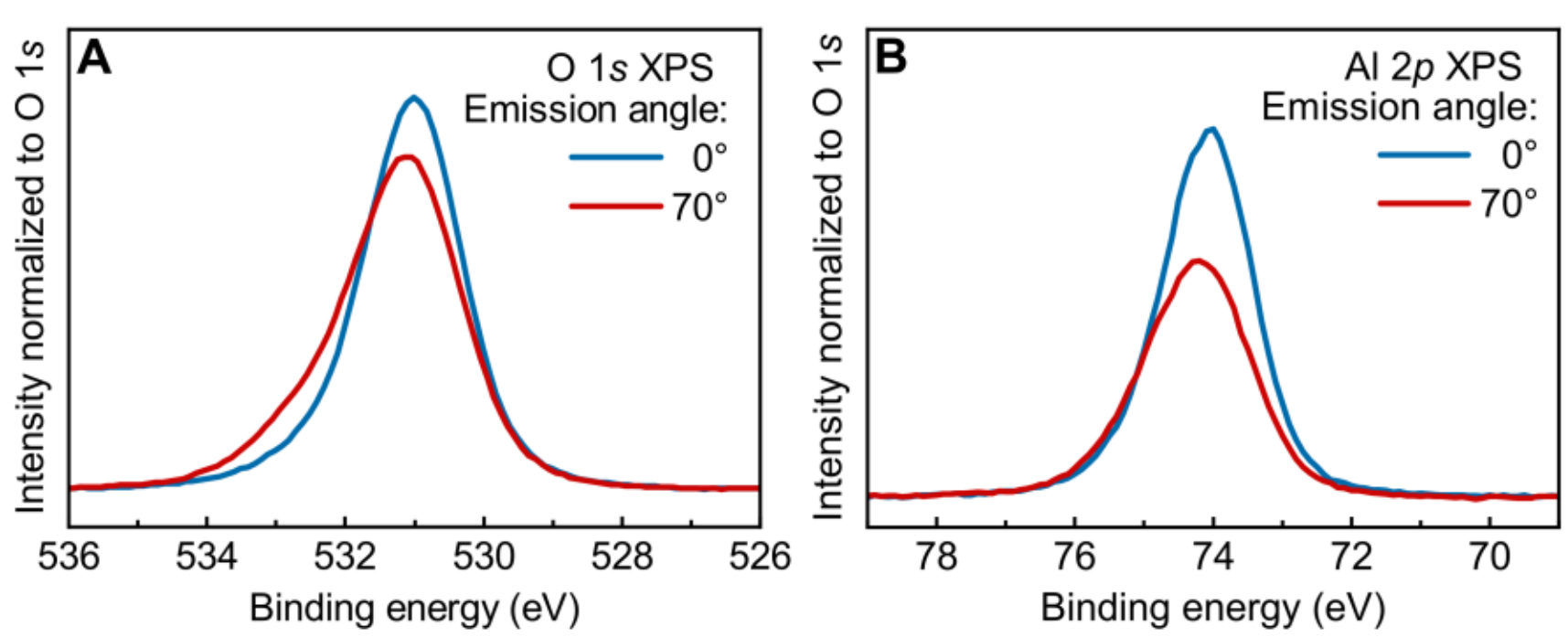

**Fig. S4. Detailed XPS spectra of the (√31 × √31)R±9° reconstructed $Al_2O_3$(0001) surface.**
(**A**) O 1*s* and (**B**) Al 2*p* spectra were collected at photoelectron emission angles of 0° (blue) and 70° (red, more surface sensitive) from the surface normal using Al Kα X-rays. The asymmetry of the 70° spectra is attributed to contributions of differently coordinated surface atoms (*15*) and local variations of the electrostatic potential. All spectra have been corrected for sample charging after subtracting a linear Shirley background. Intensities are normalized to the O 1*s* integrated area at the respective emission angle for comparison of their shapes and the relative Al content. The decrease in Al intensity at 70° emission indicates an O-terminated, not an Al-terminated surface.

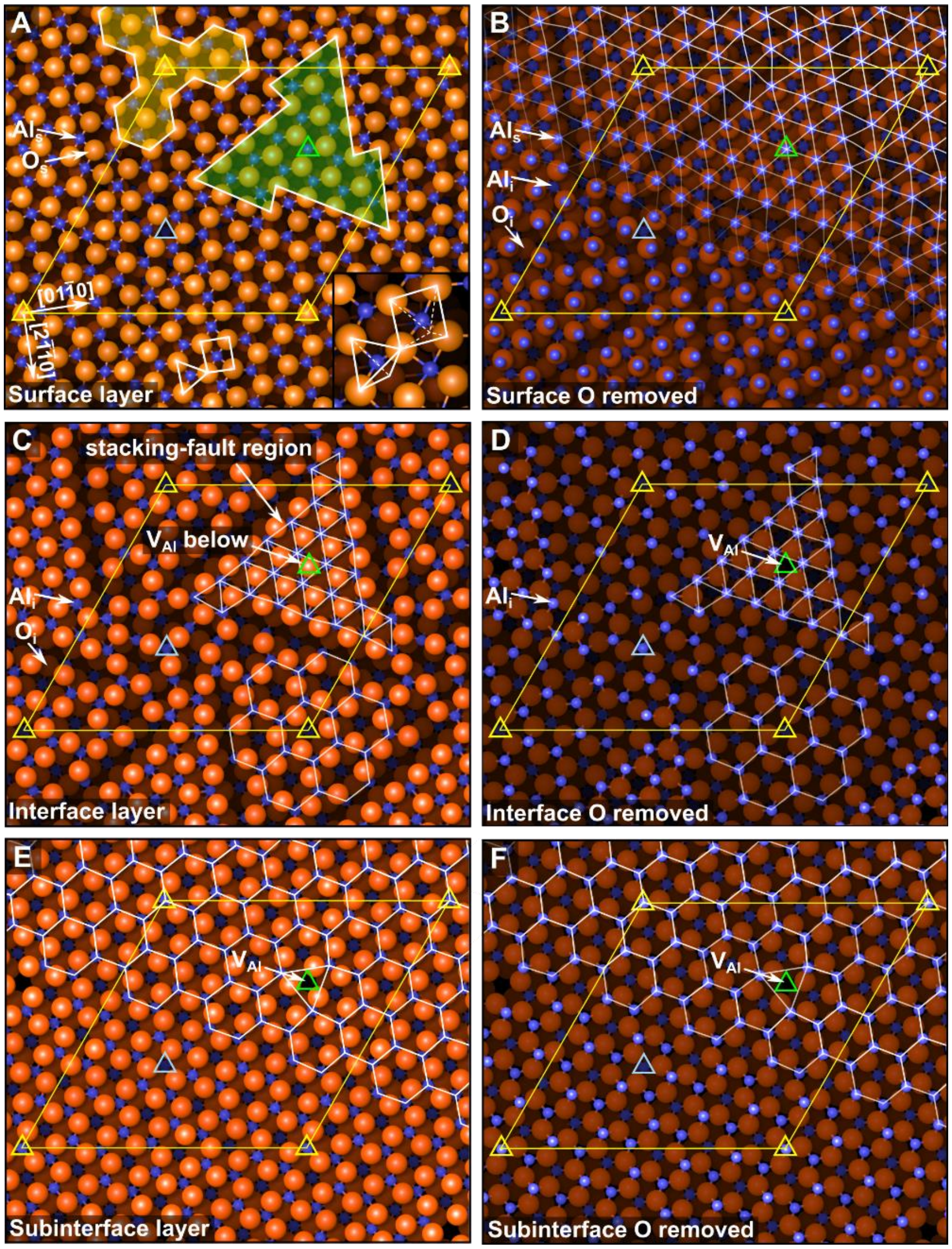

**Fig. S5. Complete structure model of the √31 reconstruction of $Al_2O_3$(0001).**

Panel (**A**) shows the surface; panels (**B**–**F**) show the layer sequence; each one was obtained by removing the topmost layer from the previous panel.

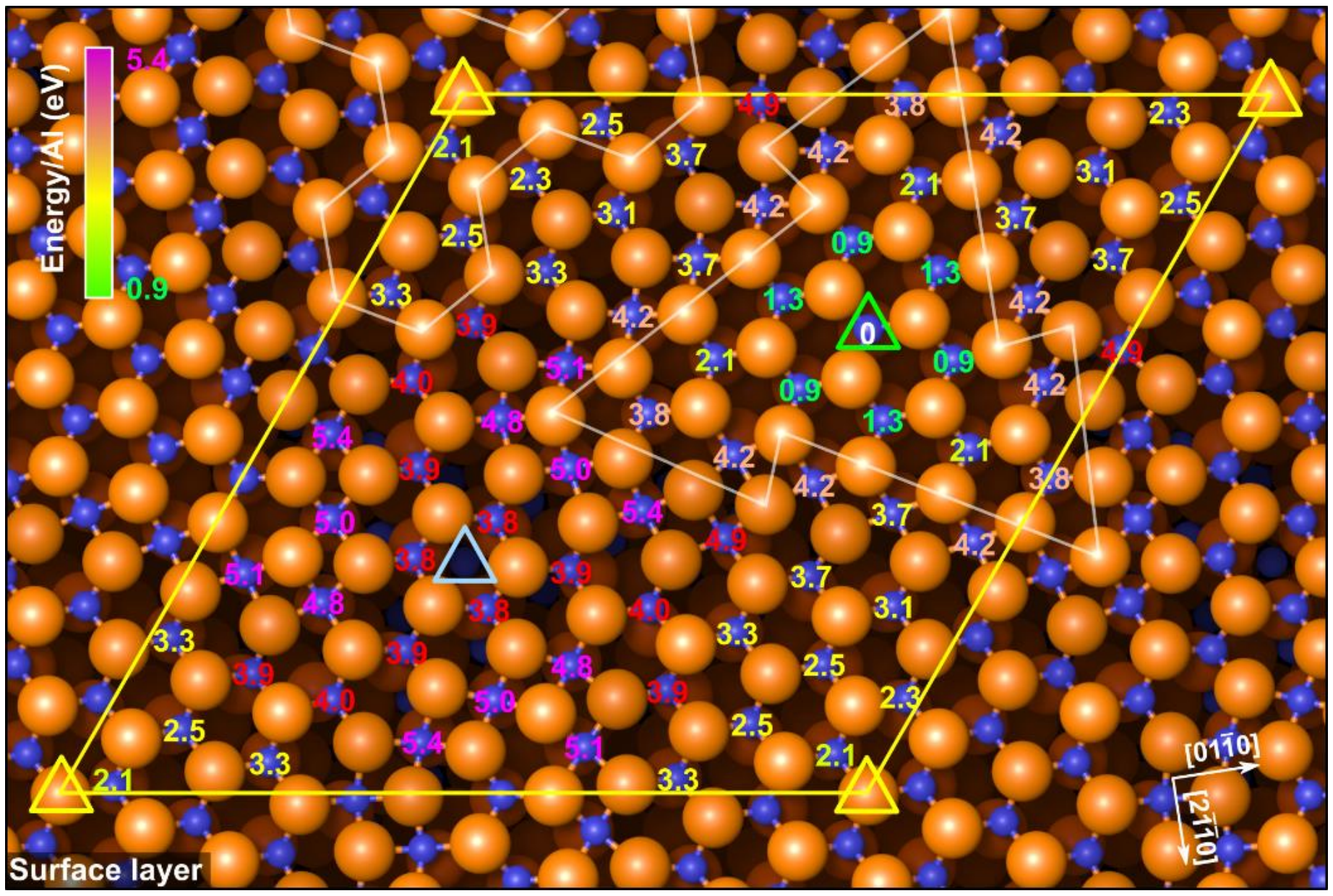

**Fig. S6. Calculated energy differences (in eV) for creating an Al vacancy at different surface sites.**

The energies were calculated in single-shot calculations (without relaxation) for an Al vacancy at various surface sites (always one surface Al atom removed). The most favorable position for creating an Al vacancy is at the symmetry center in the stacking fault region, where the $Al_i$ density is high (green triangle; the vacancy formation energy there is used as a reference, 0.0 eV). Al vacancies near the center of the stacking fault region (high local Al density) are more favorable than in the other parts of the superstructure cell. In contrast, removal of a fivefold coordinated Al atom (from a square of four $O_s$ neighbors with one $O_i$ below) is highly unfavorable ($\geq$ 3.7 eV additional energy cost per Al atom), particularly from sites far from the stacking fault, where the $Al_i$ density is low.

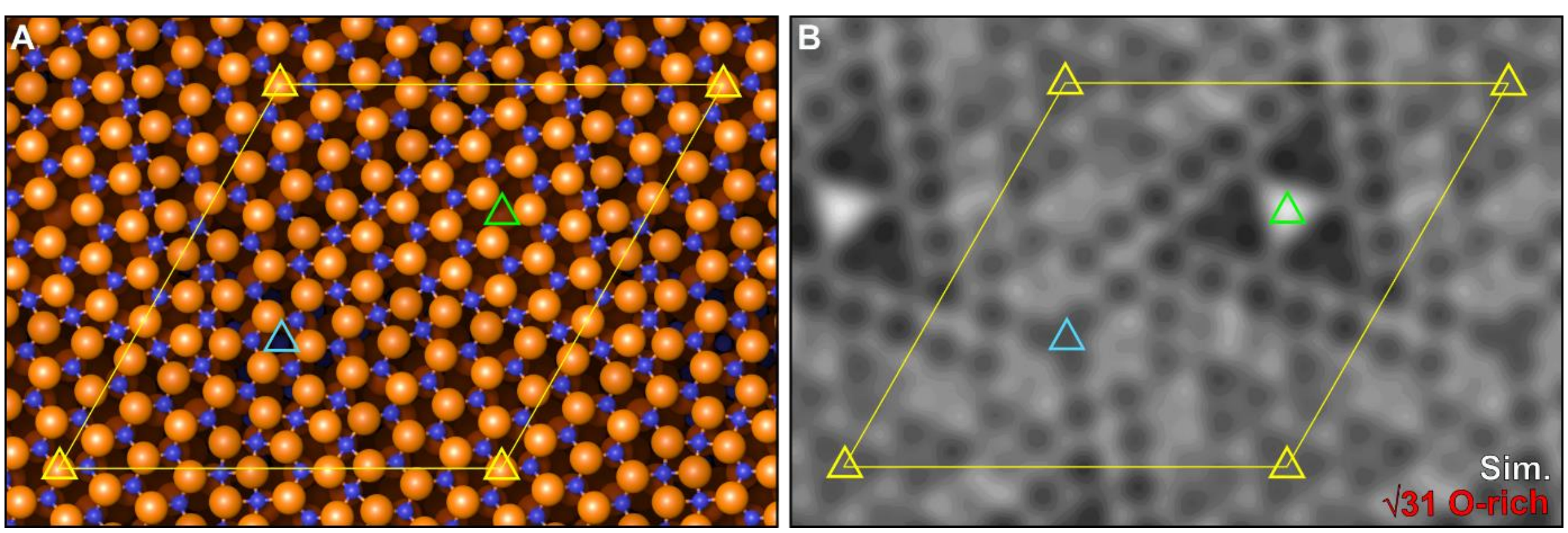

**Fig. S7. The O-rich √31 reconstructed surface with a surface Al vacancy.**

(**A**) Structure model of the surface layer, (**B**) simulated AFM image. The surface Al vacancy (inside the green triangle) causes the bright triangular appearance of the adjacent undercoordinated $O_s$ atoms (see Figs. 1 and S2).

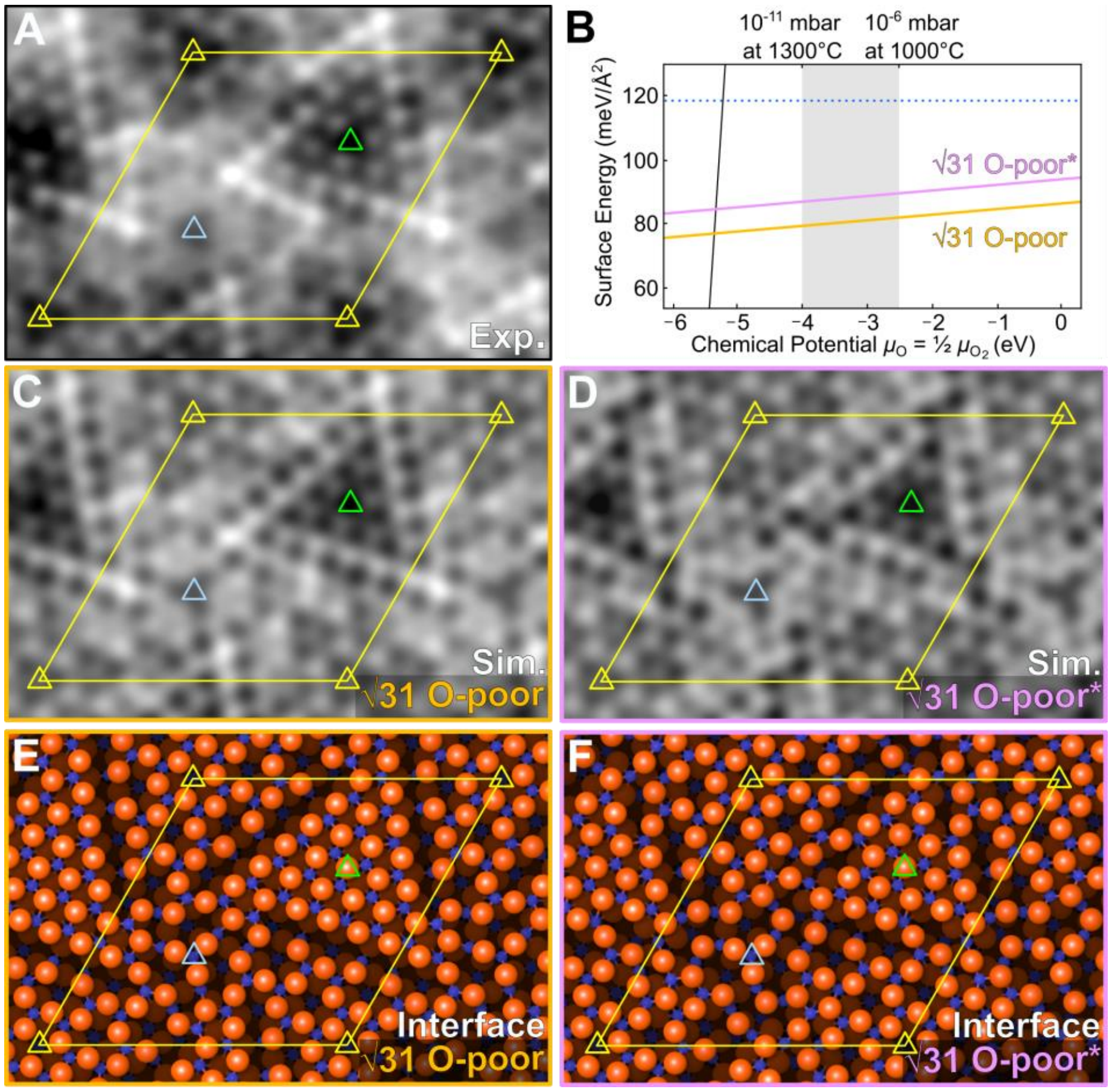

**Fig. S8. Influence of small structure variations in the interface layer on simulated AFM images.**

(**A**) Experimental, symmetry-averaged nc-AFM image of the √31 reconstructed $Al_2O_3$(0001). (**B**) *Ab initio* phase diagram of selected reconstruction models. The "√31 O-poor" (yellow line) is the energetically most stable model (Fig. 2) proposed in this work. The "√31 O-poor*" (pink line) is the second-best model with the same stoichiometry found in our MD runs. The only difference with respect to the model of Fig. 2 is a different arrangement of the atoms at the border of the stacking-fault region, resulting in a slightly higher energy. The interface structure of the two models is shown in panels (**E**) and (**F**). (**C**, **D**) Simulated AFM images obtained at a distance of 5.3 Å from the most-protruding O atom in the surface layer. The simulated AFM of the stable "√31 O-poor" model in panel (C) reproduces the features of the experimental image in (A). The noticeably worse correspondence of the less favorable "√31 O-poor*" model in panel (D) demonstrates that AFM simulations are highly sensitive to small structural changes even in subsurface layers and thus suitable to assess the correctness of a structure model.

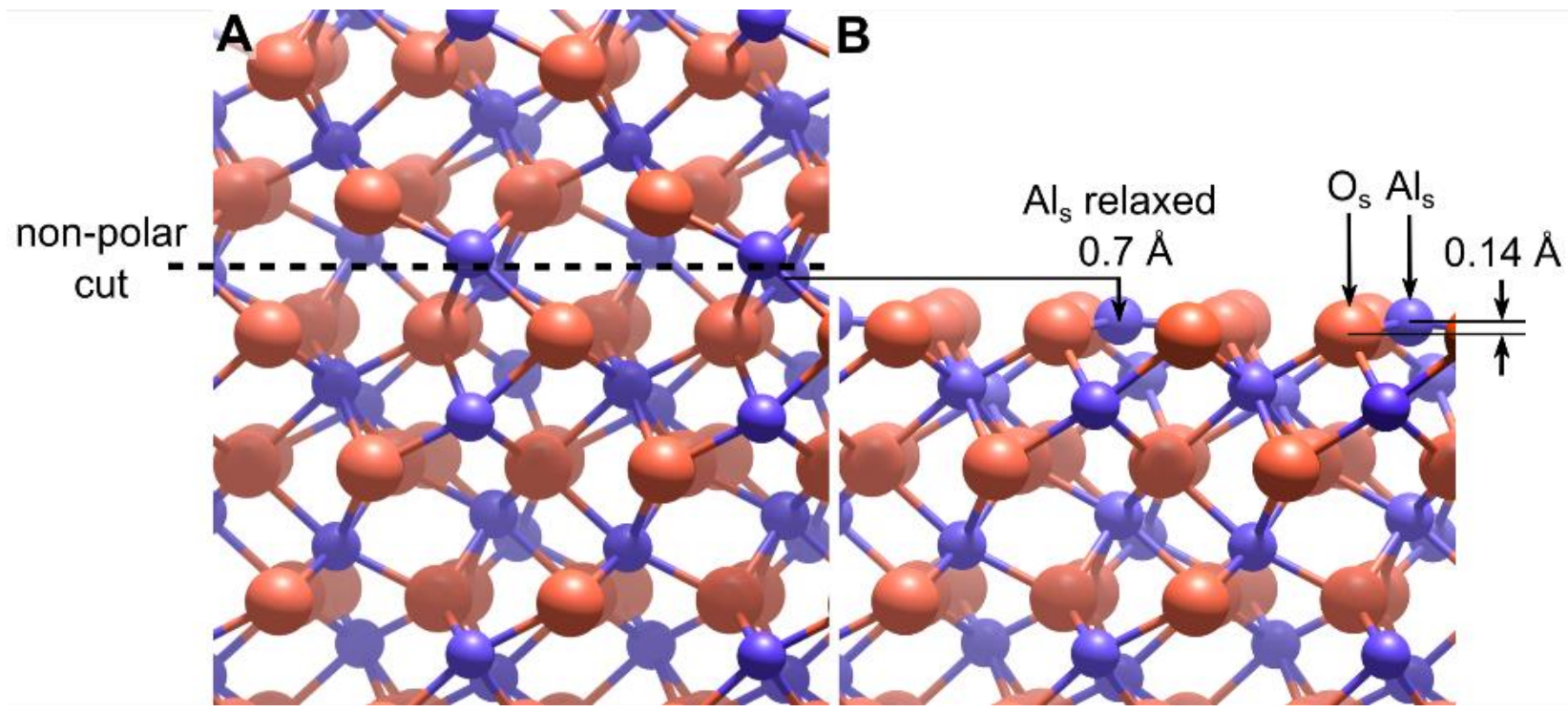

**Fig. S9. Creation of the bulk truncated (1 × 1) $Al_2O_3$(0001) surface.**

(**A**) Corundum α-$Al_2O_3$ bulk with a non-polar cutting plane along (0001) indicated by a dashed line. (**B**) DFT-relaxed (1 × 1) surface. Upon relaxation, the surface $Al_s$ atoms move inward and assume a nearly threefold-planar coordination with surface $O_s$ atoms.

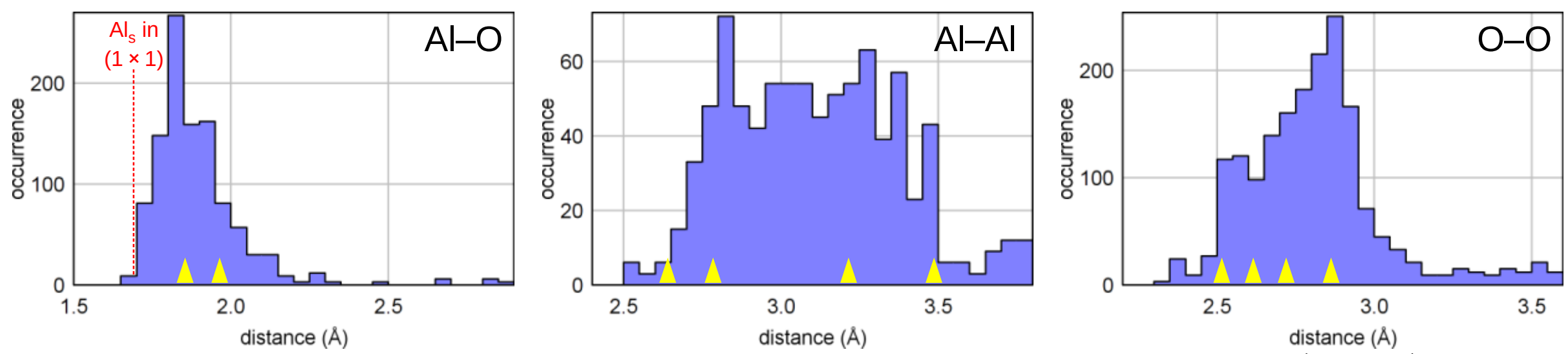

**Fig. S10. Radial distribution functions (bond length distributions) in the (√31 × √31)R±9° reconstructed $Al_2O_3$(0001).**

The graphs show the radial distribution function within the uppermost four oxygen and three aluminum layers of the model in Figs. 2 and S5. The yellow arrows indicate bond lengths in bulk corundum, and the vertical dashed line indicates the $Al_s$–O distance in the (unfavorable) (1 × 1) termination. Note that the Al–O distances in the √31 reconstruction tend to be shorter than in corundum. This is related to the shorter bond lengths of a tetrahedral vs. octahedral coordination. The distances of atoms with repulsive electrostatic interactions (Al–Al, and O–O) are comparatively large, as expected for an energetically favorable rearrangement.

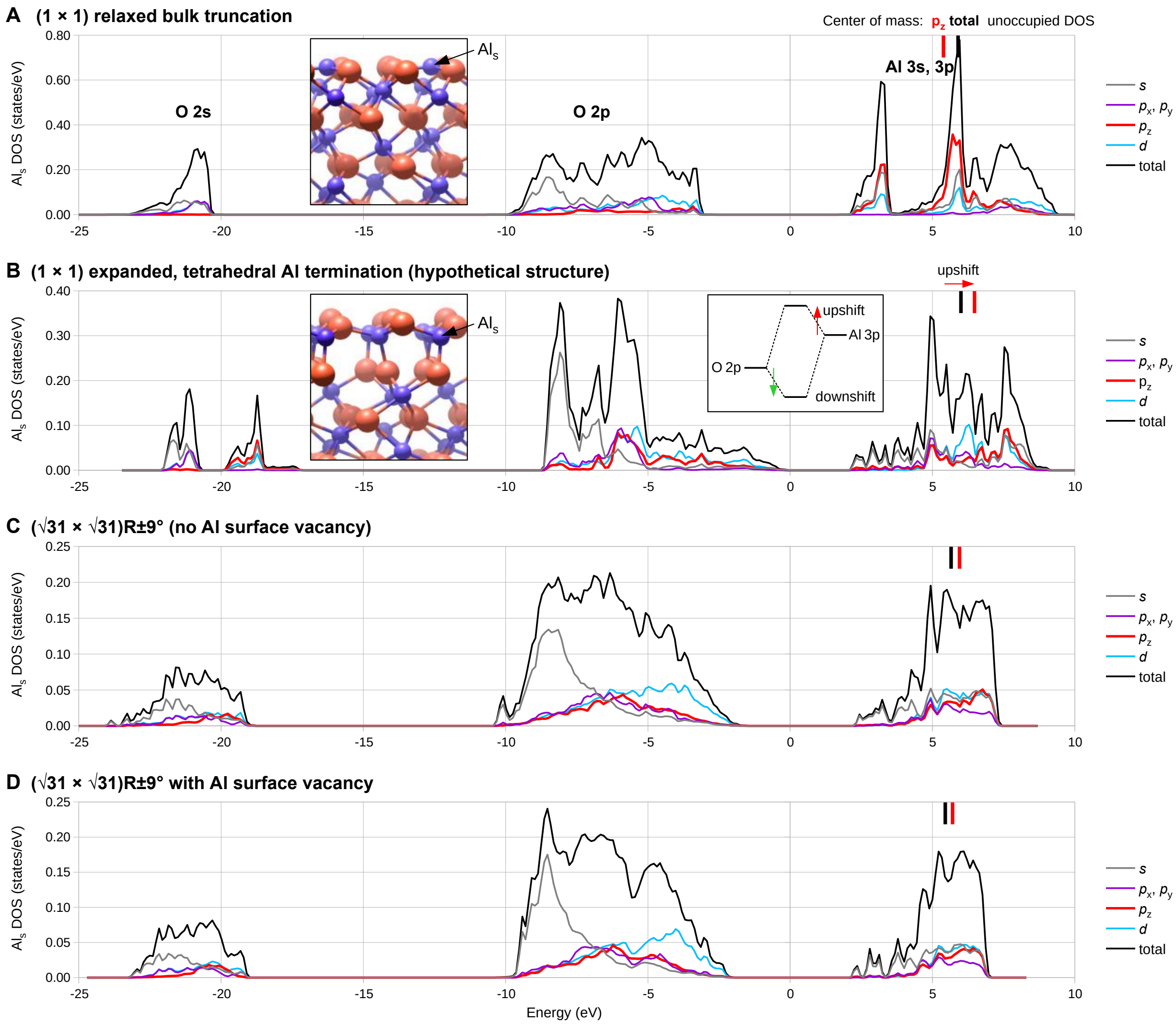

**Fig. S11. Calculated density of states of surface Al atoms.**

The local projected DOS of the uppermost Al atoms, evaluated using the projectors of the PAW method, is plotted against the energy. For the √31 cells (**C**) and (**D**), the DOS was averaged over all Al surface atoms. In the (1 × 1) relaxed bulk truncation, the $p_z$ orbitals of the $Al_s$ atoms are non-bonding. They do not hybridize with the O 2*s* and O 2*p* bands, as indicated by the almost vanishing DOS in the occupied states and the high DOS in unoccupied states between 2 and 7 eV (red line in **A**). In all the other structures, the Al $p_x$, $p_y$, and $p_z$ orbitals participate in the bonding to a similar degree and exhibit a similar degree of hybridization with the oxygen bands. (The $p_x$ and $p_y$ DOS curves are indistinguishable at this plot resolution.) The structure in (**B**) is a hypothetical, fully tetrahedrally terminated structure with (1 × 1) symmetry (left inset), with an in-plane expansion to yield an average Al–Al distance of the surface atoms similar to that in the √31 reconstruction. The ticks at the top mark the center of mass of the $p_z$ (red) and total (black) $Al_s$ unoccupied DOS. Note the upshift of the unoccupied $p_z$ energy upon bonding, as expected from a simple energy level diagram (second inset in B). Since the Fermi level in a wide-gap insulator is undefined, the energy axes have been shifted such that the (average) $Al_s$ 1*s* core level energy is the same in all four structures (**A**–**D**).

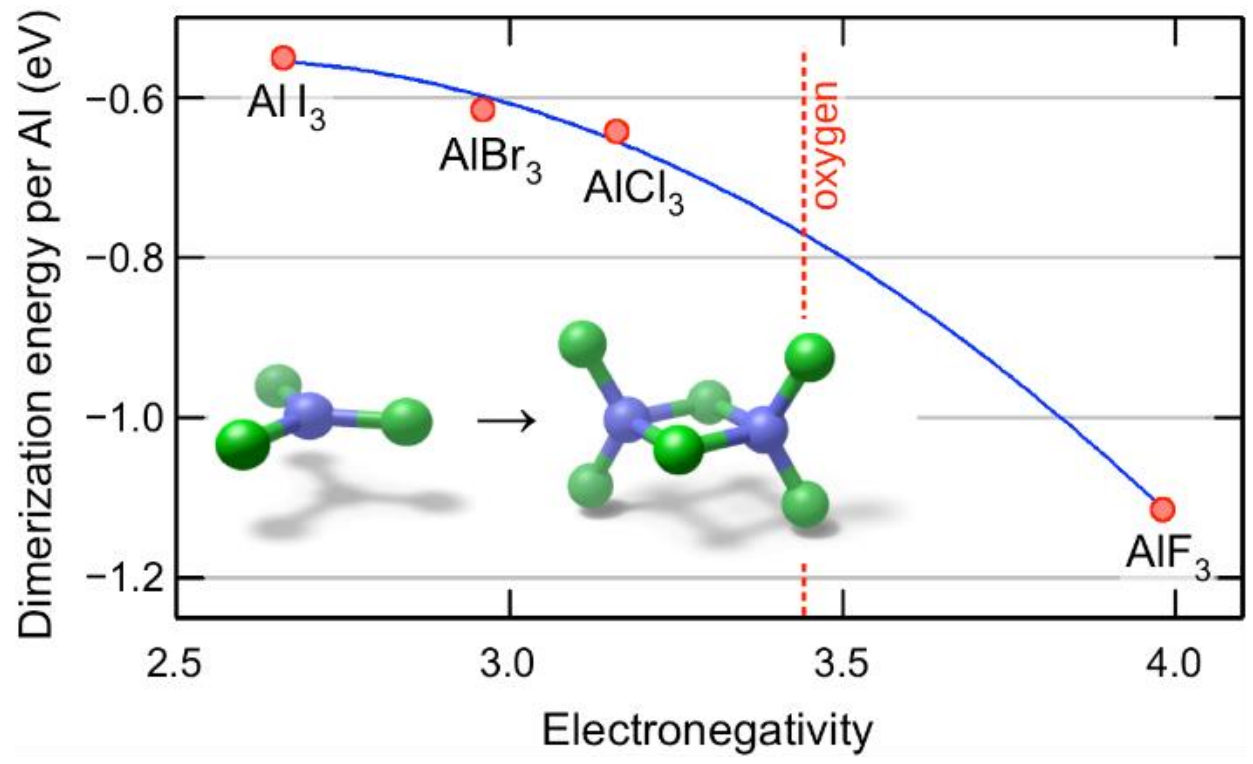

**Fig. S12. Dimerization energies of aluminum trihalides.**

The energies were taken from ref. (*16*). The blue line is a guide to the eye. The red dashed line indicates the electronegativity of oxygen. The inset shows the structures of an $AlCl_3$ monomer and $Al_2Cl_6$ dimer; the other Al(III) halides have very similar structures.